# Study on Superconducting Properties of Heavy-Fermion CeIrIn$_5$ Thin Films grown via Pulsed Laser Deposition

Ji-Hoon Kang, Jihyun Kim, Woo Seok Choi,

Sungmin Park[*], and Tuson Park[**]

*Department of Physics, Sungkyunkwan University, Suwon, 16419, Republic of Korea*

**ABSTRACT**

Herein, we report CeIrIn$_5$ thin-film growth on a MgF$_2$ (001) substrate using pulsed laser deposition technique. X-ray diffraction analysis showed that the thin films were either mainly *a*-axis-oriented (TF1) or a combination of *a*- and *c*-axis-oriented (TF2). The characteristic features of Kondo coherence temperature ($T_{coh}$) and superconductivity ($T_c$) were clearly observed, where onset of $T_c$ is 0.58 and 0.52 K, and $T_{coh}$ = 41 K and 37 K for TF1 and TF2, respectively. The temperature dependencies of the upper critical field ($H_{c2}$) of both thin films and the bulk CeIrIn$_5$ single crystal revealed a scaling behavior, indicating that the nature of unconventional superconductivity has not been changed in the thin film. The successful synthesis of CeIrIn$_5$ thin films is expected to open a new avenue for novel quantum phases that may have been difficult to explore in the bulk crystalline samples.

Keywords: CeIrIn$_5$ thin film, heavy fermions, superconductivity, metal thin film, pulsed laser deposition

[*]Corresponding author. E-mail: tp8701@skku.edu, imsuper007@gmail.com

# I. INTRODUCTION

Heavy-fermion compounds have attracted great interest because of their various novel properties, such as unconventional superconductivity, quantum criticality, and non-Fermi liquid behavior [1-3]. In strongly correlated electron systems with partially occupied $f$ orbitals, localized moments are hybridized with itinerant electrons, producing an entangled quantum state with a heavy effective mass [4]. The prototypical family, CeMIn$_5$ (M = Rh, Co, or Ir), exhibits various quantum critical phenomena that can be controlled via nonthermal parameters such as pressure, doping, or magnetic fields with relative ease because their characteristic energy scales are meV range [1, 5, 6]. Recently, dimensional tuning of CeMIn$_5$ via molecular beam epitaxy (MBE) has been successfully performed to produce superlattice structures such as CeCoIn$_5$($n$)/YbCoIn$_5$(5) (with $n$ being the number of the layer) and CeCoIn$_5$(5)/CeRhIn$_5$(5), opening a new opportunity to systematically control the novel phases of heavy-fermion systems in the reduced spatial dimension [7, 8].

CeIrIn$_5$, which crystallizes in a tetragonal structure, reaches the zero-resistance superconducting (SC) state at 1.2 K (= $T_c$) under ambient pressure [9]. However, thermodynamic studies that measured bulk properties revealed a bulk superconducting phase transition at 0.4 K, raising a question on the nature of the high-$T_c$ phase [10-11]. Recently, spatially varying superconducting domains were reported in single-crystal CeIrIn$_5$ that was cut into a tiny lamella and attached to an Al$_2$O$_3$ substrate [12]. Depending on the orientation of the transport measurements, $T_c$ varies from 200 mK to 1.2 K, suggesting that the $T_c$ difference between the transport and bulk measurements is induced from the defect-strained superconducting patches. Complex strain fields in the single crystal, which were introduced by focused ion beam (FIB) patterning, were argued as the origin of the heterogeneous electronic phases, making strain as a promising nonthermal tuning parameter of unconventional superconductivity. However, since FIB technique is not easily accessible on a bulk single crystal, thin film growth could be a more effective route for strain engineering, which is expected to deepen our understanding on the origin of the high-$T_c$ phase in CeIrIn$_5$ and other quantum phenomena that could be amplified in the reduced spatial dimension [13].

In this Letter, we report the synthesis of heavy-fermion superconducting CeIrIn$_5$ thin films using the pulsed laser deposition (PLD) technique. X-ray diffraction study revealed that two types of thin films were grown, where TF1 is mainly *a*-axis oriented, while TF2 is the mixture of *a*-and *c*-axis orientation. Electrical resistivity measurements of both thin films showed a Kondo coherence peak and a zero-resistance phase transition, which are characteristics of heavy-fermion superconductivity. The temperature dependence of the upper critical field was similar to that of the bulk single crystal, indicating that the superconductivity of the thin film was similar to that of the bulk crystal. When subjected to the magnetic field, the temperature exponent *n* of the resistivity changes from 0.95 to 1.22 and 1.15 to 1.37 for TF1 and TF2, respectively, indicating the field-induced transition from the non-Fermi liquid regime toward the conventional Fermi-liquid regime.

## II. EXPERIMENT

CeIrIn$_5$ thin films were deposited on MgF$_2$ (001) substrates using the PLD technique. The targets were prepared in-house by the arc melting method, wherein stoichiometric metal components (cerium 99.9%, iridium 99.95%, indium 99.999% purchased from Kojundo Chemical Lab) were melted together under an argon atmosphere. Contamination by oxygen was minimized using titanium metal as an oxygen absorber during the melting process. An excimer laser with a wavelength of 248 nm (LightMachinery IPEX-864) was used as a source to ablate the CeIrIn$_5$ polycrystalline target. Prior to the deposition, the MgF$_2$ (001) substrate was sequentially sonicated in acetone, ethanol, and diluted water. The base pressure of the PLD chamber was approximately

$2.5 \times 10^{-9}$ Torr and maintained at ~$10^{-8}$ Torr during deposition. Before deposition, the substrates were heated to 800 °C for 2 h to remove residual contamination on the surface. The total laser shot quantity and deposition temperature for TF1 and TF2 were 18,000 shots at 450 °C and 54,000 shots at 550 °C, respectively. The thicknesses of the films determined by field-emission scanning electron microscopy (FE-SEM) were 150 and 200 nm for TF1 and TF2, respectively. The crystal orientation of the thin films was verified using X-ray diffraction (XRD), and the lattice parameters were estimated using the FullProf software package. The thin films were categorized as TF1 and TF2 according to their crystal orientation. FE-SEM and energy-dispersive X-ray spectroscopy (EDX) were used to examine the thin-film surface morphology, thickness, and stoichiometry. Electrical transport measurements were performed using the standard four-probe technique in the temperature range of 0.3–300 K in the Physical Property Measurement System (PPMS), Oxford Teslatron, and $^3$He-cryostat system (Oxford Heliox VL).

## III. RESULTS AND DISCUSSION

Figures 1(a) and (b) show the XRD results for the $CeIrIn_5$ thin films of TF1 and TF2, respectively, deposited on the $MgF_2$ (001) substrate. For both films, $CeIrIn_5$ phase was observed with a small amount of impurity phases. In the case of TF1, the peaks (*h*00) of $CeIrIn_5$ phase were predominantly observed, revealing that the film grew along the *a*-axis. As shown in the SEM image (Fig. 1(c)), rectangular grains were formed in the TF1. However, the in-plane orientation of the rectangular layer (i.e., the *ab* or *ac* plane of $CeIrIn_5$) is not consistent with that of the $MgF_2$ substrate, indicating that the *a*-axis $CeIrIn_5$ layers were randomly oriented against the [100] $MgF_2$ substrate. This type of grain has also been observed in *c*-axis-oriented $CeCoIn_5$ films on *c*-$Al_2O_3$ and MgO substrates grown using the PLD technique [14, 15]. In the case of TF2, both *a*-axis- and

$c$-axis-oriented grains were observed with additional peaks such as (101), (220), and (312) as shown in Fig. 1(b). XRD analysis showed that the $a$- and $c$-axis lattice constants for TF1 and TF2 were $a$ = 4.666 Å and $c$ = 7.523 Å, and $a$ = 4.671 Å and $c$ = 7.513 Å, respectively, as summarized in Table I. The lattice constant, $a$, of TF1 (TF2) was slightly reduced (expanded) compared with that of a single crystal, whereas the $c$-axis constant was expanded (reduced). The expansion of or reduction in the lattice constant in comparison with that of the single crystal may be ascribed to the strain induced by the substrate.

The surface morphologies of TF1 and TF2 obtained by FE-SEM are representatively displayed in Figs. 1(c) and (d), respectively. Rectangular grains were clearly observed for TF1, which could be assigned to $CeIrIn_5$ nanocrystals with an $a$-axis orientation. Smaller and randomly shaped domains were also observed in Fig. 1(c). The surface morphology of TF2, on the other hand, suggests that grains grew with vacant regions, which spread throughout the thin film. EDX analysis shows that the average stoichiometry for the rectangular region of TF1 and the layered region of TF2 was Ce: Ir: In = 1.12: 1.00: 4.57, whereas the stoichiometry outside the rectangular region of TF1 was Ce: Ir: In = 1.10: 1.00: 3.87, indicating that cerium is slightly in excess, whereas indium is deficient for all grains and boundaries.

In the case of TF1, when the ablated ionized atoms from the target reach the substrate surface at the early growth stage, they may not be adsorbed individually on the substrate; instead, the formation of $a$-axis $CeIrIn_5$ layers may facilitate the formation of rectangular layered structures. This type of growth is also observed in $CeCoIn_5$ nano/microcrystals on $r$-$Al_2O_3$ substrates and sometimes on $MgF_2$ substrates [16]. The residual adatoms, which cannot participate in the rectangular layer formation, form a randomly oriented $CeIrIn_5$ layer with indium deficiency. The relatively high indium deficiency may be due to its low boiling point. In the case of TF2, the entire substrate was almost covered with layers, which could be associated with the increased substrate temperature and number of laser shots. The thicknesses measured from the cross-sectional FE-SEM images were approximately 150 nm and 200 nm for TF1 and TF2, respectively. The film thickness follows a linear relationship with the shot quantity, up to 18,000 shots. Afterward, the ablated atoms are consumed in forming the indium-deficient CeIrIn$_5$ layers with a higher yield, making full coverage of the layers on the substrate.

Figure 2(a) shows the normalized (divided by the 300 K value) electrical resistance for the TF1 and TF2 thin films and bulk single-crystal CeIrIn$_5$ [9]. The solid arrows mark the characteristic temperature with a local maximum—the coherence temperature ($T_{coh}$) below which conduction electrons coherently scatter off the lattice of Kondo singlets, reducing the resistance [17]. In contrast, for $T > T_{coh}$, the resistance followed a logarithmic temperature dependence, which is related to the incoherent scattering of the conduction electrons from Ce$^{3+}$ local moments [18]. When compared with the bulk single crystal (~ 66 K), the thin-film $T_{coh}$ values are shifted to lower temperatures of 41 and 37 K for TF1 and TF2, respectively.

Figure 2(b) shows the normalized resistance near the SC phase transition temperature ($T_c$), where the onset of $T_c$ (= $T_{c,on}$) was decreased from 1.32 for bulk crystal to 0.58, and 0.52 K for TF1 and TF2 thin films, respectively. The width of the SC transition $\Delta T_c$ is 0.13 K for bulk crystal, while it is 0.29 and 0.20 K for TF1 and TF2 thin films, respectively. The residual resistivity ratio (RRR) estimated by $R(T = 300\ \text{K})/R(T_{c,on})$, as summarized in Table II, is significantly reduced from 46.8 for bulk crystal to 5.63 and 5.43 for TF1 and TF2 thin films, respectively, indicating that the increase in $\Delta T_c$ as well as the suppression in $T_c$ is influenced by disorder in thin films [19].

The electrical resistances of TF1 and TF2 are plotted as a function of temperature near $T_c$ for several magnetic fields applied along the MgF$_2$ [001] direction in Figs. 3(a) and (b), respectively. The dependence of $T_{c,on}$ on the magnetic field is plotted in Fig. 3(c), where the upper critical fields ($H_{c2}$) of TF1, TF2, and the single crystal for $H$ applied perpendicular and parallel to the $c$-axis are described by black circles, red squares, blue up triangles, and green down triangles, respectively. The dashed lines represent the least-squares fittings of the phenomenological expression $H_{c2}(T) = H_{c2}(0)[1 - (T/T_{c,on})^2]^\beta$, where the fitting parameters are summarized in Table II. Figure 3(d) shows the scaling behavior of the reduced upper critical field ($H_{c2} / H_{c2}(T = 0 \text{ K})$) as a function of the reduced temperature ($T / T_c (\mu_0 H = 0 \text{ kOe})$). All datasets of CeIrIn$_5$ thin films and single crystals are collapsed on top of each other where the value of $\beta$ was 0.73 (see the red dash-dotted line), indicating that the nature of superconductivity in thin films is similar to that of bulk crystals.

Figures 4(a) and (b) show the electrical resistivity of TF1 and TF2, respectively, where 4 T (red circles) applied along the MgF$_2$ [001] direction was sufficient to suppress the SC phase. Similar to the bulk crystal, the resistance of thin films approximately followed the linear-in-T dependence: the temperature exponent ($n$) obtained from the least-square fits of $\rho(T) = \rho_0 + AT^n$ was 0.95 ± 0.01 and 1.15 ± 0.01 for TF1 and TF2, respectively. As the magnetic field was increased from 0 to 4 T, $n$ increased from 0.95 to 1.22 and from 1.15 to 1.37 for TF1 and TF2, respectively. These results indicate that the non-Fermi liquid behavior observed at 0 T is associated with a quantum critical point (QCP), and the applied magnetic field drives the system away from the QCP [6, 21].

We note that the zero-resistance state of both TF1 and TF2 thin films is achieved at around 300 mK, which is significantly suppressed from that of the high-$T_c$ phase ($T_c$ = 1.2 K), but is very close to that determined from thermodynamic measurements. Lattice parameters determined from X-ray diffraction analysis (see Table I) indicate that both TF1 and TF2 films are under strain, possibly from the lattice mismatch with the substrate. The fact that both TF1 and TF2 shows lack of high-$T_c$ phase seems at odds with the scenario that the high-$T_c$ phase of CeIrIn$_5$ arises from the defect-strained patches. However, residual resistivity ratio (RRR) is 5.63 and 5.34 for TF1 and TF2 thin films, respectively, which is significantly lower than that of single crystalline sample (= 46.8), indicating that disorder is an important factor in determining the SC phase transition temperature. Further work on the growth of high-quality epitaxial thin film and the systematic control of disorder could be critical to properly understanding the mystery of the high-$T_c$ phase in CeIrIn$_5$.

## IV. CONCLUSION

In summary, we report the first successful deposition of CeIrIn$_5$ thin films on MgF$_2$ (001) substrates using the PLD technique. The coherence temperatures as well as the SC transition temperatures of the thin films were lower than the bulk single-crystal case, which could be related to the increase in the degree of disorder of the thin films. When plotted against the reduced temperature, the normalized upper critical fields of the thin films and bulk crystals were collapsed into a single curve, indicating that the nature of the superconductivity of thin films is similar to that of bulk single crystals. These results indicate that the PLD technique can be utilized as an alternative deposition method for the growth of heavy-fermion superconducting thin films. Further efforts to grow high-quality epitaxial thin films are expected to shed light on the evolution of the Kondo coherence behavior as well as unconventional superconductivity in the heavy-fermion superconductor family.

## ACKNOWLEDGMENTS

This work was supported by a National Research Foundation (NRF) of Korea grant, funded by the Korean Ministry of Science and ICT (No. 2021R1A2C2010925). S Park, J. Kim, and J. –H. Kang were partially supported by the Basic Science Research Program through the National Research Foundation of Korea (NRF), funded by the Ministry of Education (NRF-2016R1A6A3A11935214 and NRF-2018R1D1A1B07051040). W. S. Choi was supported by NRF- 2021R1A2C2011340.

**FIGURES**

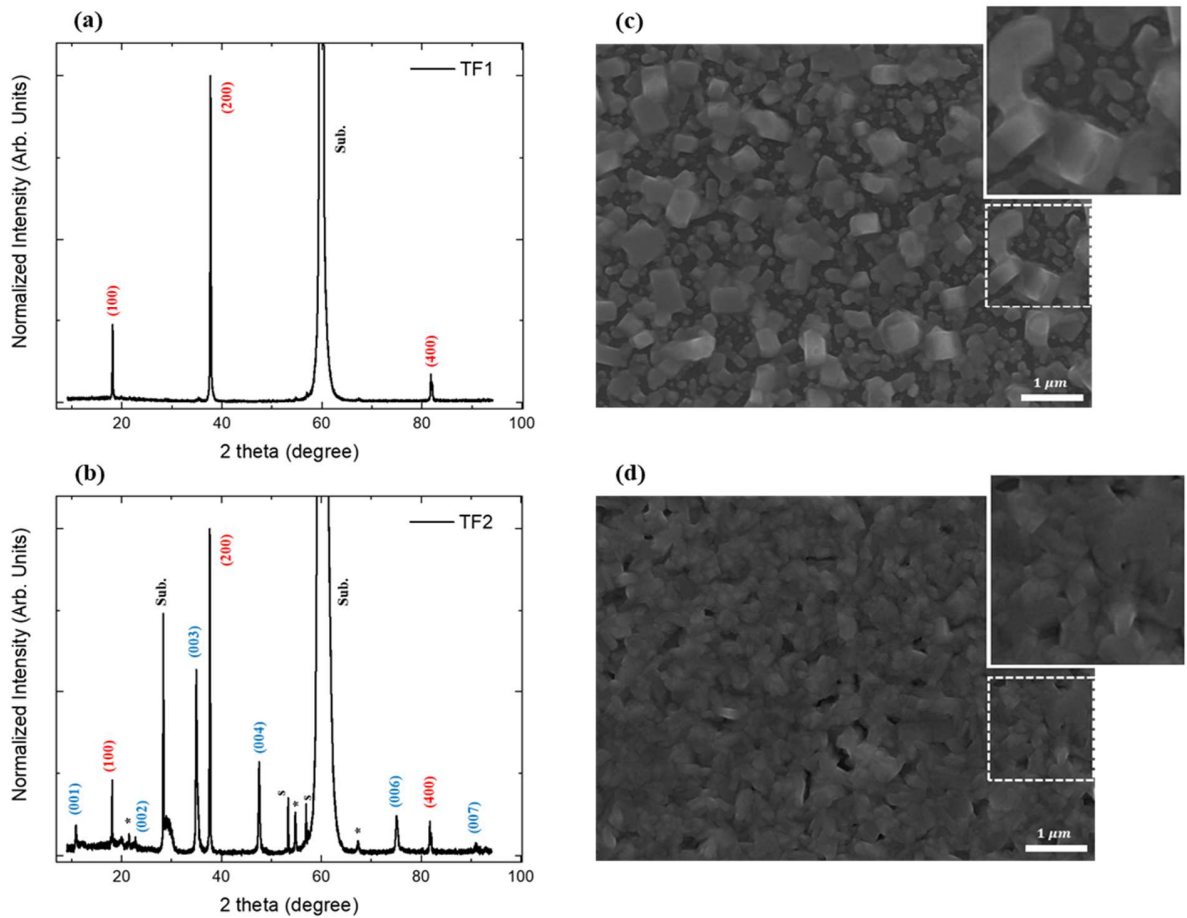

FIG. 1. X-ray diffraction and FE-SEM results of TF1 and TF2 deposited on a MgF$_2$ (001) substrate. The *a*- and *c*-axis orientations of TF1 and TF2 are denoted by the red and blue miller indices in (a) and (b), respectively. TF1 shows mainly *a*-axis oriented thin-film growth, whereas TF2 shows biaxial (combination of *a*- and *c*-axis orientation) growth. The different surface morphologies of TF1 and TF2, which were probed by FE-SEM, are shown in (c) and (d), respectively. The inset in the SEM displays the magnified image of the region indicated by the white dotted square. The thickness was measured to be 150 and 200 nm for TF1 and TF2, respectively.

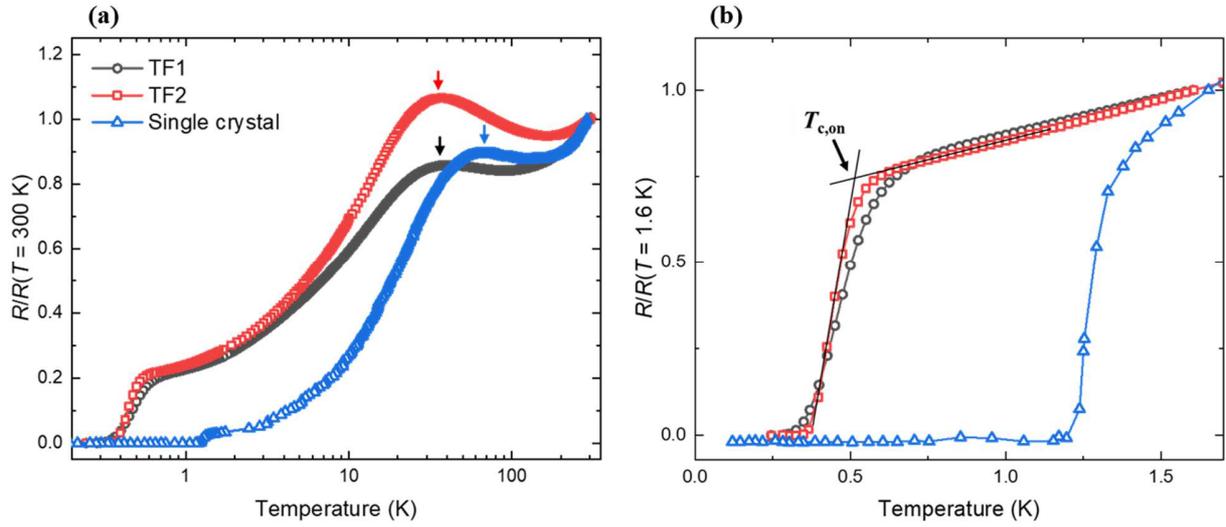

FIG. 2. (a) Electrical resistances of TF1, TF2, and bulk crystal, which are normalized by the resistance at 300 K, are plotted as a function of temperature. Solid arrows mark the coherence temperature ($T_{coh}$), where the resistance reveals a peak. (b) Electrical resistance is magnified near the SC phase transition. When compared with the bulk crystal, the onset temperatures of SC transition ($T_{c,on}$) of the thin films were suppressed. The arrow marks the criteria for the $T_{c,on}$ of this study.

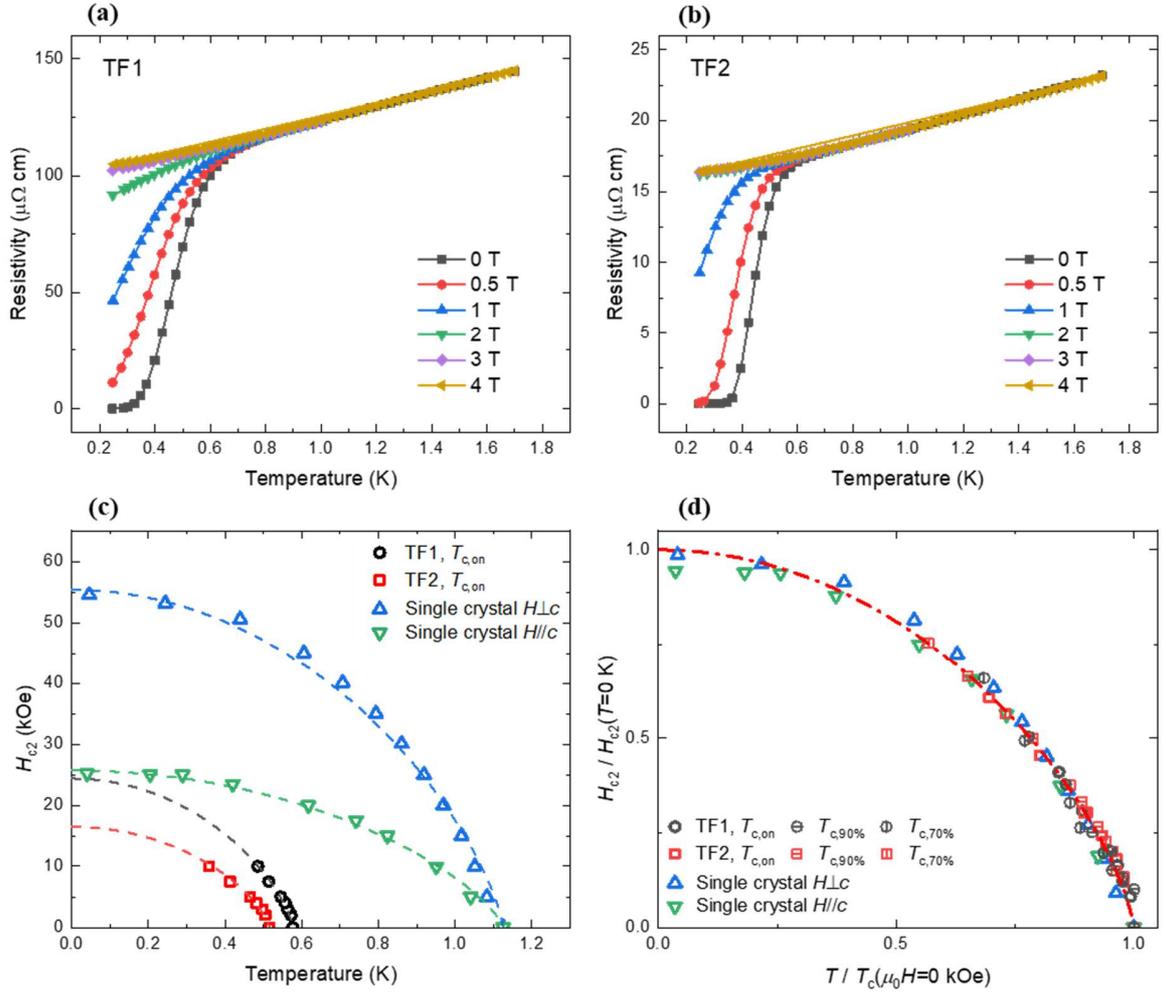

FIG. 3. Electrical resistances of TF1 and TF2 are plotted against temperature for several magnetic field in (a) and (b), respectively. (c) The upper critical field determined from $T_{c,on}$ is plotted as a function of temperature for TF1 (black circles), TF2 (red squares), and bulk single crystal for $H \perp c$ (blue up triangles) and $H // c$ (green down triangles). The dashed lines are the best results obtained from the least-squares fittings of the phenomenological expression $H_{c2}(T) = H_{c2}(0)[1 - (T/T_{c,on})^2]^\beta$. (d) Reduced upper critical field as a function of reduced temperature for CeIrIn$_5$ single crystals and thin films. Upper critical fields determined from $T_{c,90\%}$ and $T_{c,70\%}$ are additionally plotted. The scaling behavior (red dash-dotted line) showed that the temperature dependence of the reduced upper critical field was independent of the criterion used to determine $T_c$.

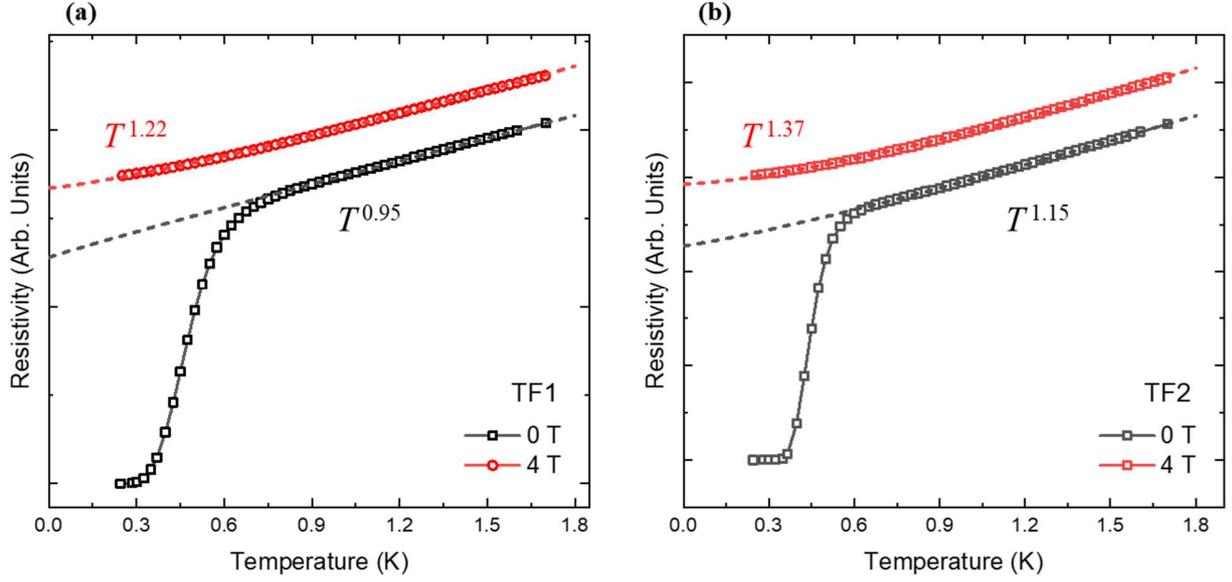

**FIG. 4.** Temperature dependence of resistance under magnetic fields of 0 (black squares) and 4 T (red circles) for (a) TF1 and (b) TF2. The 4 T resistivity curve is shifted for clarity. The dashed lines show best results from the least-squares fitting of $\rho(T) = \rho_0 + AT^n$. The temperature exponent of the CeIrIn$_5$ thin films increased with application of magnetic field.

TABLE I. Summary of the surface morphology and structural properties of single-crystal and thin-film CeIrIn$_5$.

|  | Morphology | $a$ (Å) | $c$ (Å) | $c/a$ |
|---|---|---|---|---|
| Single Crystal[a] |  | 4.668 | 7.515 | 1.610 |
| TF1 | Island | 4.666 (-0.043%) | 7.523 (+0.106%) | 1.612 |
| TF2 | Grained Layer | 4.671 (+0.064%) | 7.513 (-0.027%) | 1.608 |

[a] Reference [9]

TABLE II. Summary of superconducting transition temperature, upper critical field (fitting parameter), coherence temperature, and residual resistivity ratio (RRR) of single crystal and thin-film $CeIrIn_5$.

| | $T_{c,on}$ (K) | $T_{c,zero}$ (K) | $\Delta T_c$ (K) | $H_{c2}(0)$ (kOe) ($\beta$) | $T_{coh}$ (K) | RRR |
|---|---|---|---|---|---|---|
| Single Crystal[a] | 1.32 | 1.19 | 0.13 | ($H \perp c$) 55.4 (0.735)<br>($H // c$) 25.7 (0.737) | 66.6 | 46.8 |
| TF1 | 0.58 | 0.29 | 0.29 | 24.4 (0.712) | 40.9 | 5.63 |
| TF2 | 0.52 | 0.32 | 0.20 | 16.5 (0.708) | 36.9 | 5.43 |

[a] Reference [20]